\documentclass[12pt]{article}
\usepackage{graphicx,times,latexsym,epsfig,rotate,pslatex}

\newcommand{\bit}{\begin{itemize}}
\newcommand{\eit}{\end{itemize}}

\newcommand{\fnaldocheader}[2]{
}

\fnaldocheader{FERMILAB-TM-2248}{May 2004}

\begin{document}

\title{Radiation Shielding Calculations for MuCool Test Area at Fermilab}


\author{I.~Rakhno, C.~Johnstone$^{*}$ \\
\normalsize \itshape{University of Illinois at Urbana-Champaign, Urbana, IL 61801}\\
\normalsize {\em $^{*}$Fermilab, P.O. Box 500, Batavia, IL 60510}
       }

\date{\today}

\maketitle

\begin{abstract}

The MuCool Test Area (MTA) is an intense primary beam facility derived directly from the Fermilab Linac to
test heat deposition and other technical concerns associated with the
liquid hydrogen targets being developed for cooling intense muon beams.
In this shielding study the
results of Monte Carlo radiation shielding calculations performed using the \textsc{MARS14} code 
for the MuCool Test Area and
including the downstream portion of the target hall and berm around it, access pit,
service building, and parking lot
are presented and discussed within the context of the proposed MTA experimental configuration.

\end{abstract}

\section{Introduction}

The MTA facility is being designed to test targets and other muon cooling apparatus using the intense
Fermilab Linac beam.
The requested intensity of the proton beam for the MTA is essentially full Linac capability \cite{bib:MuCool}, or
 $1.3\times 10^{13}$ protons
per pulse at a 15 Hz repetition rate and an energy of 400 MeV.
This intensity represents a factor of two beyond the current safety envelope of Fermilab Linac.
If it is later determined the safety envelope cannot practically be exceeded, 
the reduced intensity is still acceptable and sufficient to test the MuCool targets and apparatus.

This extremely high intensity implies careful investigation into and application of 
proper shielding materials and configuration in order
to satisfy the following two requirements: 
(i) to reduce the instantaneous dose rate outside of the experimental enclosure to prescribed levels appropriate 
for the area considered;
(ii) to ensure the civil construction of the hall is capable of additional shielding and, further,
that the weight of the shielding is  
commensurate with the loading specifications of the enclosure, notably the ceiling.

The radiation shielding calculations
for the MuCool experimental enclosure were performed with the \textsc{MARS14} \cite{bib:mars} code for both normal operation and
accidental beam loss. Normalization is per $2\times 10^{14}$ protons per second unless otherwise stated.
 Various shielding options were explored in detail and the final, most effective, and, therefore, minimal, 
shielding configurations are presented.
The possible factor of two reduction as indicated above
does not effectively alter the shielding conclusions and requirements established in this document. 
Gerrymandering or reduction of the shielding can only be accomplished unless the
intensity is reduced by an order of magnitude from full capability. 
Further, additional shielding above the
enclosure (beyond the present berm level) will be required unless the intensity is reduced by another
order of magnitude, or two orders of magnitude down from Linac capability.

\section{Geometry Model}

\begin{figure}[htb!]
\vspace{1mm}
\hspace{12mm}
\includegraphics[width=3.9in,keepaspectratio] {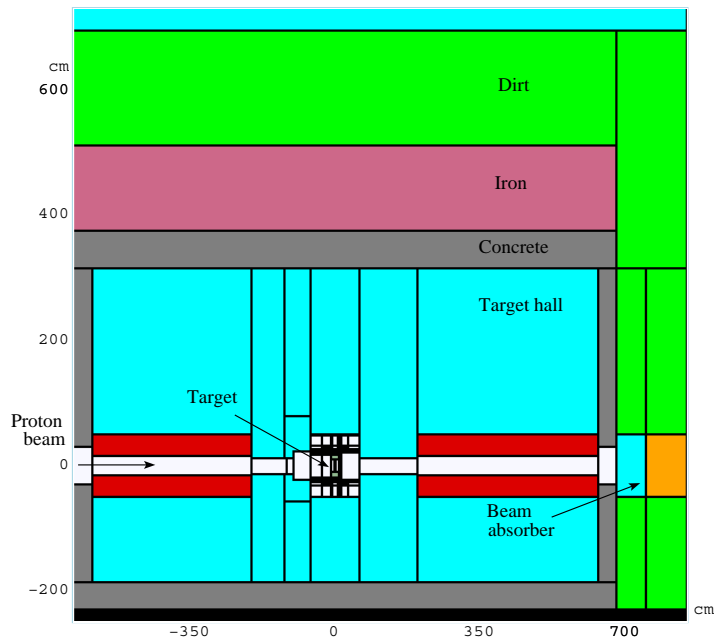}
\vspace{-10mm} \hspace{39mm} \caption {An elevation view of the
\textsc{MARS14} model of the MTA.} \label{target_hall_XZ}
\end{figure}

\noindent
A Cartesian coordinate system is established in which a
three-dimensional geometry model of the enclosure, target and beamline components,
and exterior shielding is described. The $X$- and $Z$-axes are shown
in Fig.~\ref{target_hall_XZ}, with the $Y$-axis being directed
toward a reader to complete the right-handed coordinate system.
Thus, the positive direction for the $X$-axis is upward while for
the $Z$-axis it follows the direction of the beam, \emph{i.e.} downstream. The
origin of the coordinate system, $(0,0,0)$, is chosen at the
geometrical center of the target (see Fig.~\ref{target_hall_XZ}).
Previous studies have addressed activation and loss conditions on
experimental components in the enclosure.  
However, for the purposes of shielding against normal operation, the target
and beam absorber are the only significant sources for radiation outside
the enclosure.  Further, it has been established, as will be discussed,
that normal operation
represents the severest radiological control problems, not accident conditions
as this is a target hall.

As for the color scheme employed to denote materials in the
geometry model, the following convention applies to any system:
white, black, light blue, green, and grey colors correspond to
vacuum, black hole (artificial material used in \textsc{MARS}
modeling that absorbs 100\% of incoming radiation), air, soil, and
regular concrete, respectively. (The meaning of the other colors can
vary depending on materials used in the system under consideration.)
It should be taken into account also that boundaries between different
regions are shown with black lines. When the resolution of the figure
is inadequate, small regions sometimes are not distinguishable and
appear as black regions.

\subsection{Target hall}

A cross-sectional slice of the three-dimensional calculation model of the MTA is presented in detail in Fig.~\ref{target_hall_XZ}.
It consists of the downstream 40 feet of the target hall, the target itself with associated windows and cryogenics,
the beamline, the beam absorber, and the surrounding shielding.
The shielding layers of iron and dirt shown in the Figure represent only one of many shielding options
and configurations which were modeled. (The current enclosure shielding contains only 11 feet of berm.)  
The upstream portion of the target hall, which is a pre-cast concrete enclosure, is
approximately 30 feet in length, 10 feet in width, and is not considered in the model. 
The upstream portion stretches from the shield-block wall under the hatch to the ``step'', where floor level drops
more than 2 feet in elevation from 738 feet 7 inches to 736 feet 6 inches.
Calculation of the optimal shielding thickness and composition for the experimental enclosure
is the main goal of the study and discussed in the
following sections.

For the purposes of a thorough and more complete study,
we considered two target models: a copper disk 1 cm in thickness in addition to the liquid hydrogen as designed for MuCool.
Taking into account the data on the proton interaction lengths presented in Table~\ref{proton_lengths}, the models
with the copper disk and 21-cm liquid hydrogen absorber with two 200-$\mu$m aluminum windows correspond to 
10\% and 2\% of the
proton total interaction length, respectively.  Thus the model with the copper disk enables us to perform the radiation shielding
assessment for a more general dose rate as would be expected for operations which involve targets 
other than liquid hydrogen, or
more windows, or evacuation of liquid hydrogen from the target, or even alternate ``thicker-window'' 
designs for liquid hydrogen absorbers.  
(Liquid hydrogen is effective in reducing the dose rates at the top of the berm by about 30\% 
for 400 MeV incident protons as compared with an evacuated target.
However, one cannot rely
on liquid hydrogen always being present in the beam in this facility.)
The soil considered in the study is supposed to be compacted one with the density 
characteristic of the Fermilab site, \emph{i.e.} 2.24 g/cm$^3$.

\vspace{-2mm}
\begin{table}[htb!]
\caption{Proton total and inelastic interaction lengths (cm) at kinetic energy of 400 MeV.}
\begin{center}
\begin{tabular} {|c|c|c|c|}
\hline
                     &   Liquid hydrogen   &   Aluminum   &   Copper   \\
\hline
  $\lambda_{tot}$    &  \ \ 911            &  28.6        &   10.3     \\
  $\lambda_{inel}$   &  1108               &  40.8        &   15.6     \\
\hline
\end{tabular}
\end{center}
\label{proton_lengths}
\end{table}

\subsection{Labyrinth and access pit}

\begin{figure}[htb!]
\vspace{4mm} \hspace{6mm}
\centering\epsfig{figure=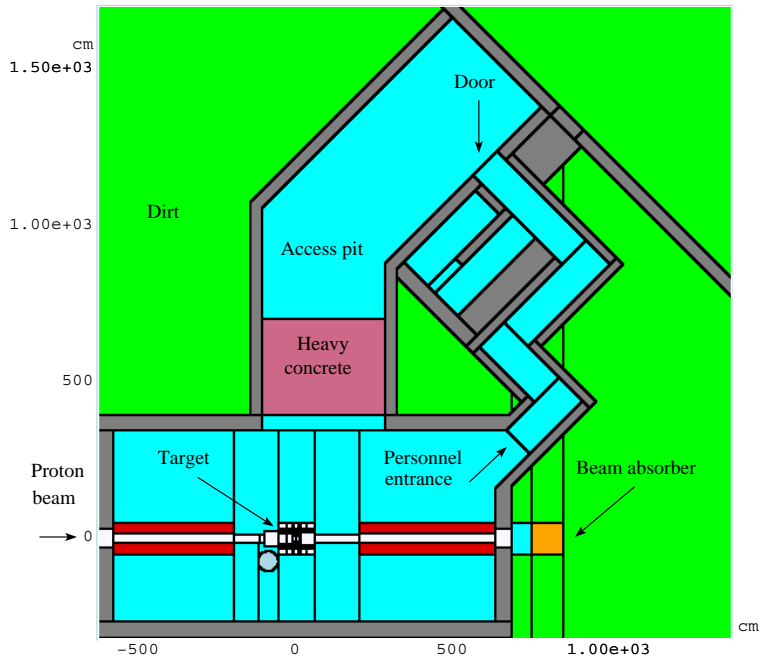,width=0.66\linewidth}
\vspace{5mm} \hspace{42mm} \caption {A plan view of the MARS14
model of the MTA at lower level.} \label{mta_lower_level}
\end{figure}

Contributions to the computed dose levels were taken from
a model of the lower level of the area, which extends over the region of $-190$ cm $\leq X \leq -20$ cm, and
which is presented in Fig.~\ref{mta_lower_level} at $X=0$ to show simultaneously both the labyrinth 
personnel entrance and target
(see Fig.~\ref{target_hall_XZ}). This range in the $X-$region was chosen to represent an average human height.
All the essential components of the MTA were included in the model, in particular,
the labyrinth between the access pit and target hall. It will be shown below that, because of relatively high
dose levels, one must consider the access pit as a ``Radiation Area''. Therefore, unauthorized access
to the pit must be prohibited when the proton beam is on.

\subsection{Penetrations and service building}

\begin{figure}[htb!]
\vspace{3mm}
\hspace{6mm}
\centering\epsfig{figure=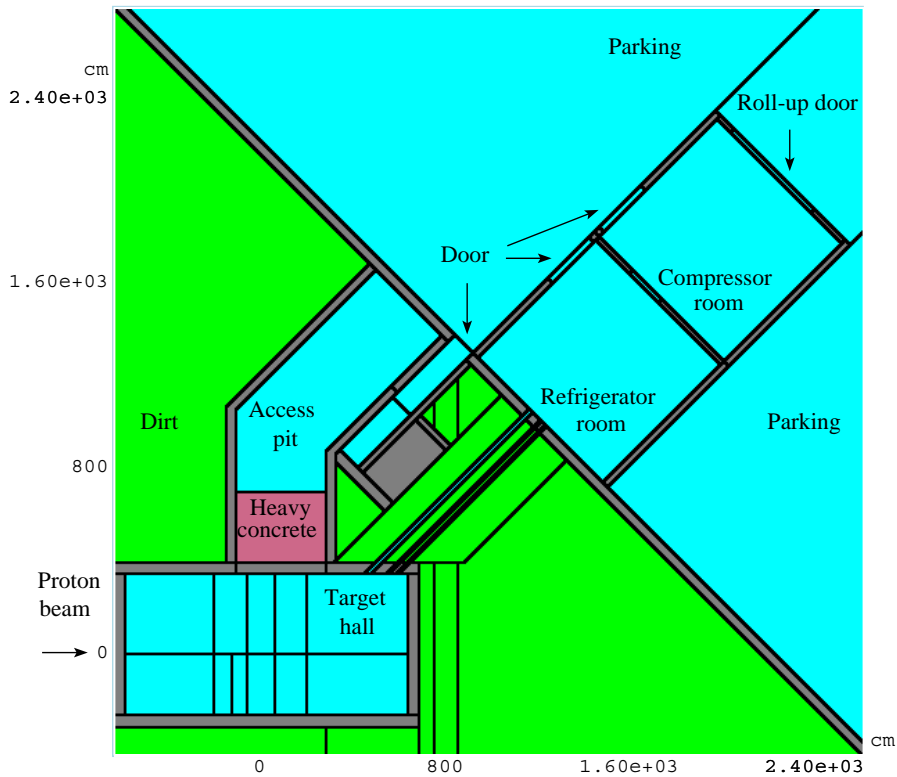,width=0.78\linewidth}
\vspace{8mm}
\hspace{25mm}
\caption
{A plan view of the MARS14 model of the MTA at upper level.}
\label{mta_upper_level}
\end{figure}

The model of the upper level of the area, $90$ cm $\leq X \leq
260$ cm (see Fig.~\ref{target_hall_XZ}), is shown in
Fig.~\ref{mta_upper_level}. There is also an internal door located
approximately in the middle of the wall between the refrigerator
and compressor rooms . The model includes, in particular, six
penetrations (channels) between the target hall and refrigerator
room. Two of the penetrations (10$^{\prime\prime}$ and
8$^{\prime\prime}$ in diameter) are designated for helium transfer
lines while the other four ones (4$^{\prime\prime}$ in diameter
each) are reserved for future use. In this model it is
assumed that the latter four penetrations are filled with air.

This level is of major concern because it includes parking lots
near Fermilab booster tower and, in fact, is open to general
public. It will be shown below that the most significant source of
radiation at this level is comprised of high-energy (100--300 MeV)
neutrons delivered through the penetrations.

\clearpage

\section{Calculation Results}

\subsection{Berm around the target hall}

\subsubsection{Accidental beam loss}

The scenario which describes the worst accidental beam loss occurs when the errant proton beam hits the beamline at Z=-280 cm
(see Fig.~\ref{target_hall_XZ}) with a deflection angle of 50 mrad upward.  The beam can acquire such a deflection
due to tuning or malfunction of upstream magnets.  
Within the framework of this scenario, the calculated highest prompt
dose in the shielding is observed right above the target assembly, \emph{i.e.} near Z=0 (see Fig.~\ref{target_hall_XZ}).
This scenario describes the worst possible case (when modeling 8-GeV proton beam accidents
at Fermilab booster, a deflection angle of about 1 mrad is the standard considered~\cite{Mokhov}).
If mis-steered beam hits the beamline downstream of the target
assembly, additional or increased thickness in shielding layers come into play 
(either beamline components, beam absorber shielding, lower angles through the shielding, or all of these) 
thus providing lower prompt
dose above the berm when compared to our assumed worst scenario.

\begin{figure}[htb!]
\vspace{1mm}
\centering\epsfig{figure=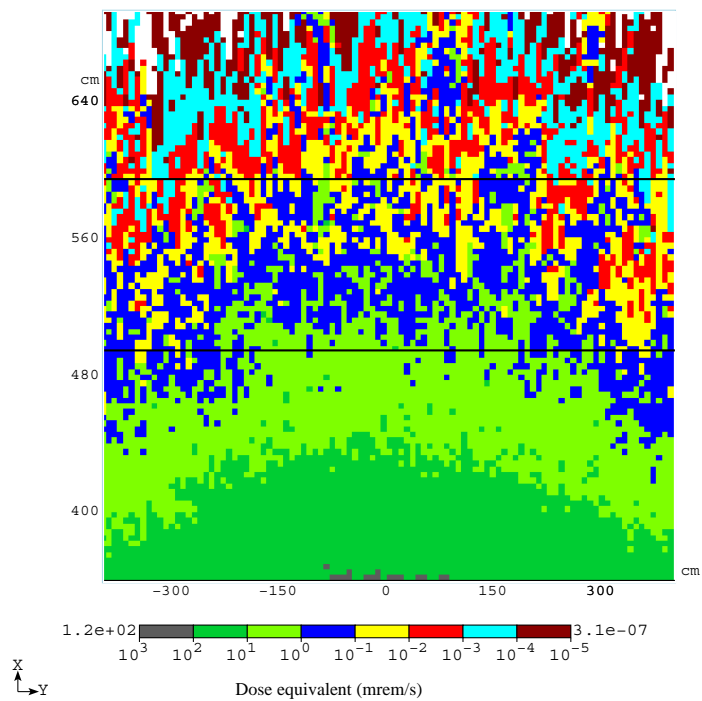,width=0.61\linewidth}
\vspace{-3mm}
\hspace{15mm}
\caption
{Calculated distribution of prompt dose equivalent in the dirt shielding above the MTA target hall and near Z=0 due to the accidental beam loss.
The calculation was performed with the liquid hydrogen absorber as the target.}
\label{accidental_beam_loss}
\end{figure}

The calculated distribution of prompt dose in the shielding above the target hall is shown in 
Fig.~\ref{accidental_beam_loss}. The numbers on the left and right of the color bar correspond to
the highest and lowest value, respectively, presented in the two-dimensional histogram (the regions with values 
outside of the \emph{current} limits are shown with white color).  
To estimate dose rate on the top of the
berm one can use a dose attenuation curve calculated previously for 400-MeV protons in similar shielding \cite{bib:dose_attenuation_curve}
starting from a region with well-defined dose.
In this way one can determine that the dose rate on the top of the initially proposed 11-feet dirt shielding is about
0.016 mrem/s in general. The target itself is of secondary importance for the dose above the shielding 
after the missteered beam strikes the beamline. 

With only 11$^{\prime}$ of shielding in place, an active system (chipmunks) protects the Linac enclosure
against accident conditions and could, in principle, be applied to the experimental hall enclosure. However,
the integrated accidental dose rate is comparable to the levels experienced during normal operation
for every level of shielding (since beam strikes a target). An active system can not protect against normal
operation, therefore, the passive shielding necessary for normal operation is sufficient to shield against
achievable accident conditions. Active interlocks are not required. (Active interlocks will be explored upstream
of the experimental hall up to the point of Linac extraction since no target sources are involved.)


\subsubsection{Normal operation}

\begin{figure}[htb!]
\vspace{0mm}
\hspace{7mm}
\centering\epsfig{figure=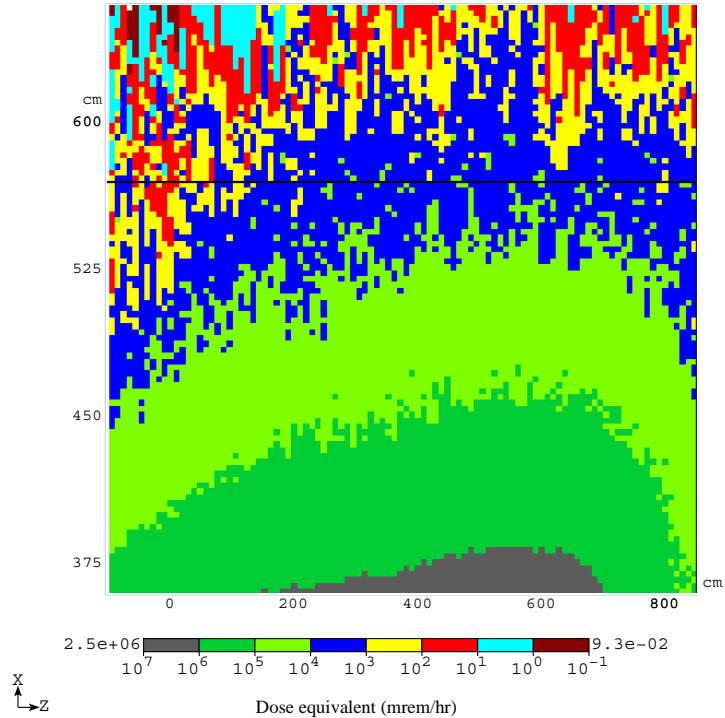,width=0.63\linewidth}
\vspace{-1mm}
\hspace{19mm}
\caption
{Calculated distribution of prompt dose equivalent in the dirt shielding above the MTA target hall at normal operation.
The calculation was performed with the 1-cm thick copper disk as the target.}
\label{dose_at_normal_operation}
\end{figure}

\noindent
A calculated distribution of prompt dose in the shielding above the target hall for normal operation 
is shown in Fig.~\ref{dose_at_normal_operation}.
Similar two-dimensional distributions were obtained also for
various iron-dirt and BMCN-dirt compositions.  The BMCN stands
here for a high-density concrete (3.64 g/cm$^3$) that contains, in
particular, 55\% of iron by weight while for regular concrete the
number is 1.4\%. To determine the amount of required shielding,
the above-mentioned dose attenuation curve of reference
\cite{bib:dose_attenuation_curve} can be used again.  For a more
convenient analysis we used also a simple expression describing
dose attenuation in a thick shielding sandwich:

\begin{equation}
\label{math/1}
    Attenuation Factor = e^{- \bigl(\frac{x_1}{\alpha_1} + \frac{x_2}{\alpha_2} \bigr)},
\end{equation}

\vspace{3mm}
\noindent
where $x_1$ and $x_2$ are thicknesses of the first and second material, respectively, and
$\alpha_1$ and $\alpha_2$ are attenuation lenghts for these materials.  Deviations of a
real attenuation law from such a pure exponential one
can be neglected for thick shieldings~\cite{bib:dose_attenuation_curve}.

\begin{figure}[htb!]
\vspace{1mm}
\centering\epsfig{figure=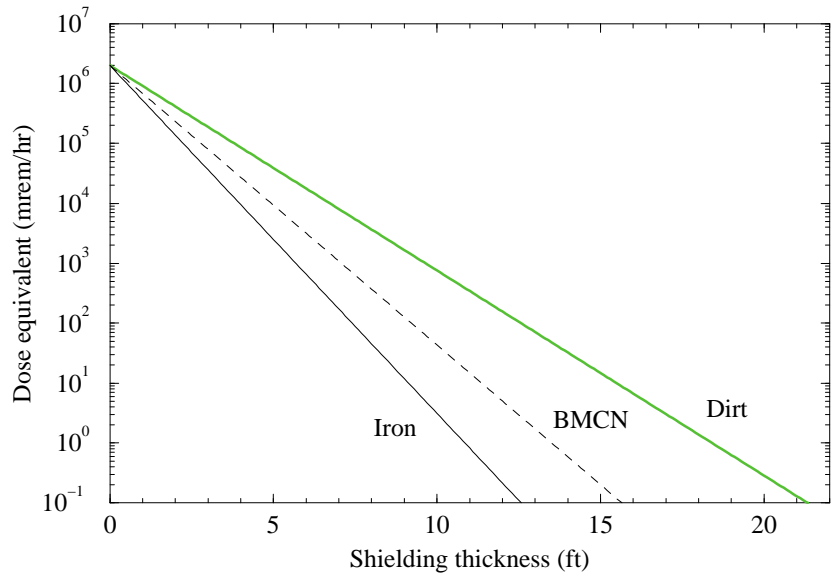,width=0.74\linewidth}
\vspace{0mm}
\caption{Calculated dose rate on the top of the MTA berm at normal operation \emph{vs} shielding thickness
for a general target (copper disk 1 cm in thickness).}
\label{pure_dirt}
\end{figure}

First of all, using the expression (\ref{math/1}) we have calculated the dependence
of the dose rate on the top of the berm
on thickness of the uniform shielding (see Fig.~\ref{pure_dirt}). This distribution is useful as a starting
and comparison point in the analysis of the shielding requirements.
The data verifies that a minimum of 16.4$^{\prime}$ of compacted dirt is needed to prevent the
top of the berm from being defined as a radiation area (which means dose rate above 5 mrem/hr)
and the 19$^{\prime}$ convention is the recommended level.

\begin{figure}[htb!]
\vspace{1mm} \noindent
\begin{minipage}[t]{0.48\linewidth}
\centering\epsfig{figure=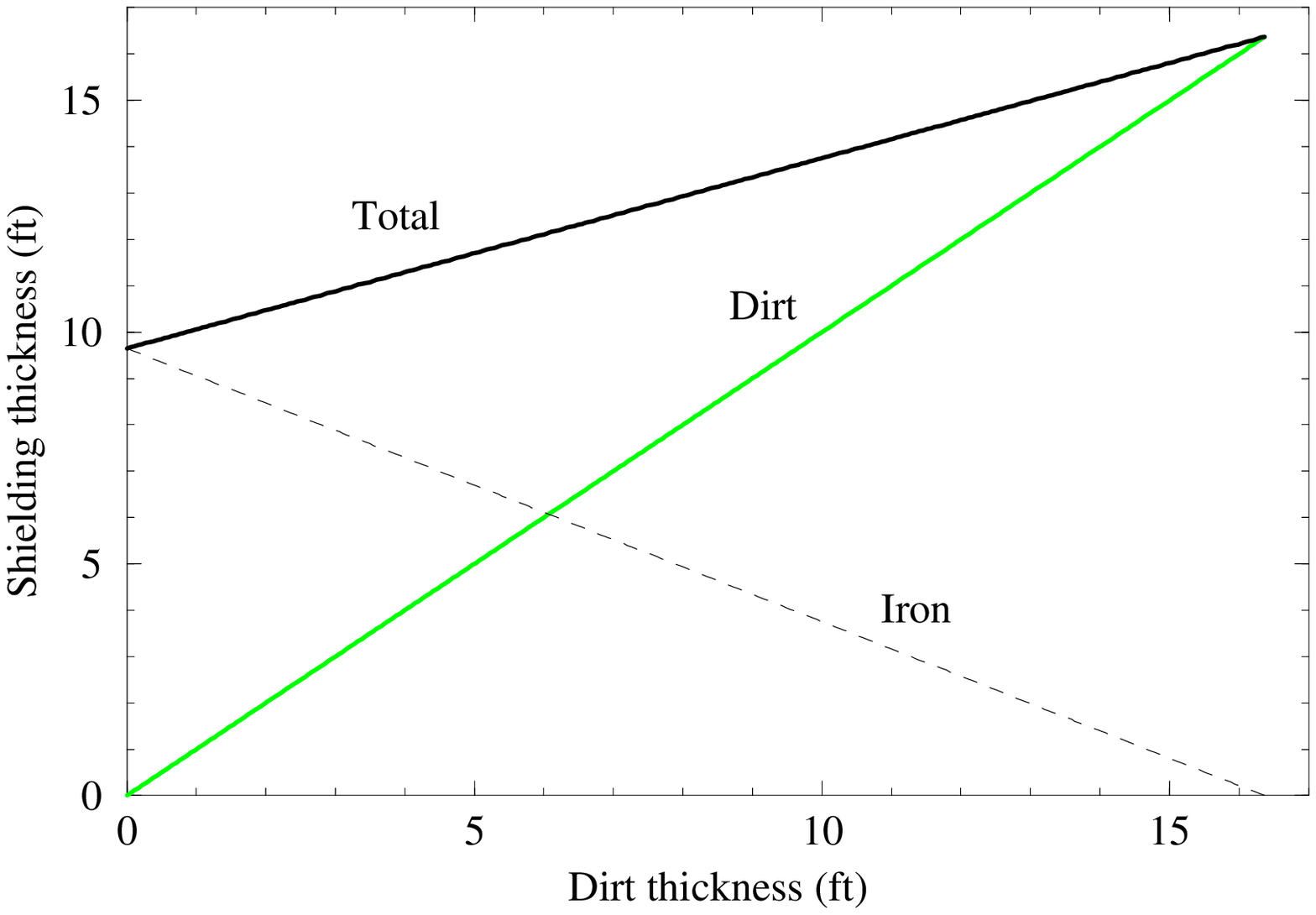,width=\linewidth}
\end{minipage}
\hfill
\begin{minipage}[t]{0.48\linewidth}
\centering\epsfig{figure=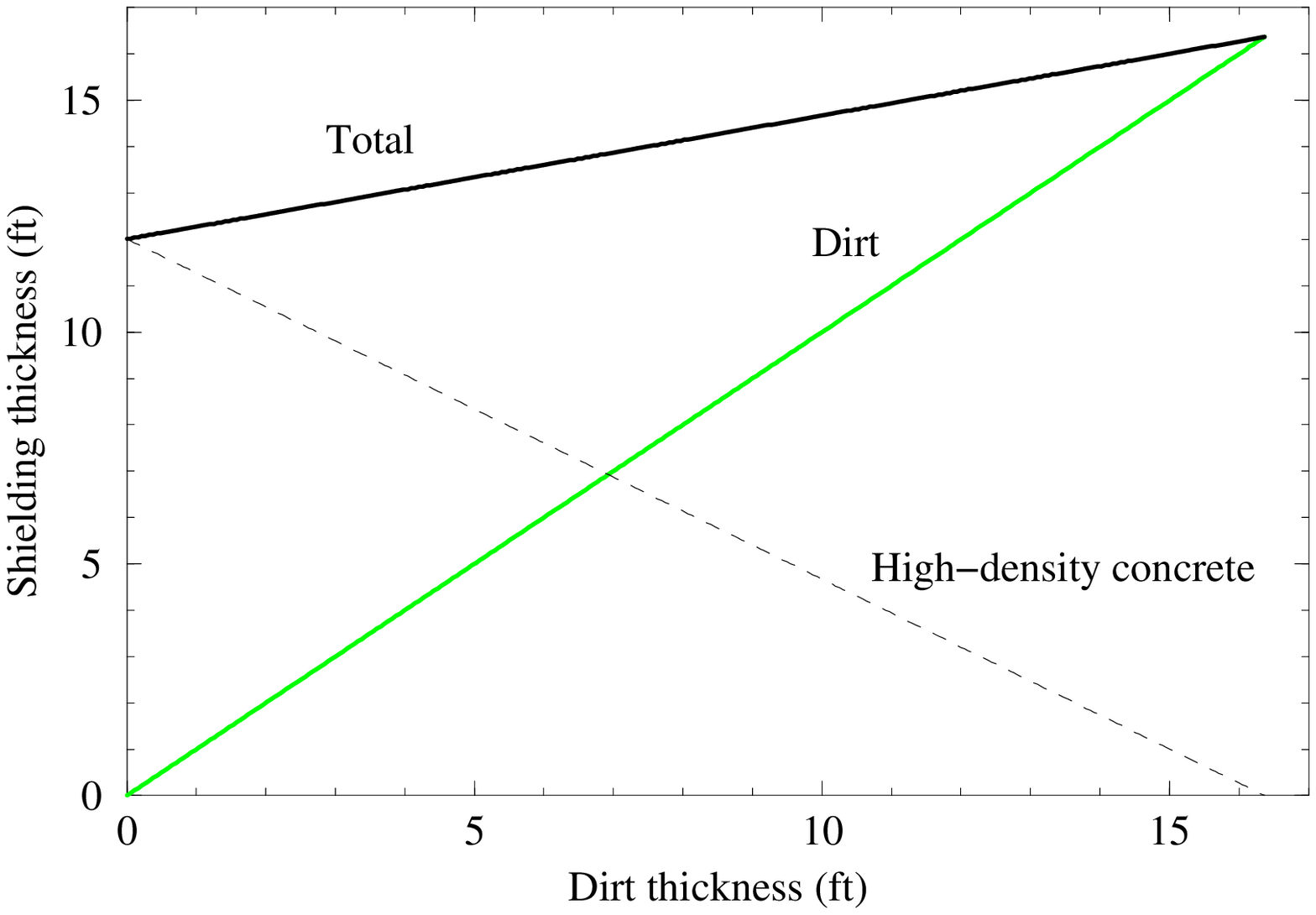,width=\linewidth}
\end{minipage}
\vspace{2mm}
\caption{Calculated shielding compositions for iron-dirt (left) and BMCN-dirt (right) sandwiches which provide the dose level of
5.0 mrem/hr on the top of the MTA shielding at normal operation.}
\label{compositions}
\end{figure}

Using the expression (\ref{math/1}) we have also calculated shielding compositions which provide a required attenuation factor
for iron-dirt
and BMCN-dirt sandwiches (see Fig.~\ref{compositions}).  The calculated attenuation lengths $\alpha$ for the dirt, high-density
concrete, and iron
were equal to 38.7, 28.4, and 22.8 cm, respectively.
The difference observed between the two predictions for pure dirt shielding (0.6 feet or, in other words, 3\%)
is due to the approximation associated with the expression (\ref{math/1}) and calculated attenuation lengths $\alpha$.
One can see from the distributions that, taking into account the weight of the shielding, compositions of the high-density
concrete and dirt look preferable when compared to iron-dirt sandwiches.
This is best illustrated by comparing weight with shielding
effectiveness.  Iron is 7.87 g/cm$^3$ compared to 3.64 g/cm$^3$ for heavy concrete, or a 2.2 to 1 weight ratio,
but their relative shielding effectiveness is only 1.2 to 1 for the same volume, respectively. 
This effect is due to the fact that the radiation  propagating through the berm consists mostly of secondary
nucleons generated in the target in inelastic nuclear collisions. The average energy of such secondaries is in
a few MeV region. At this low an energy, there is no advantage in using pure iron as the shielding material.
Materials containing light nuclei are the most effective.
  
With a load capacity of 19$^{\prime}$ equivalent of dirt, and assuming the top layer must be 2$^{\prime}$ of berm 
to stop thermal
neutrons, the enclosure can only support 17 equivalent feet of dirt by weight.
This corresponds to 10.5$^{\prime}$ of heavy concrete and only 5$^{\prime}$ of iron.   With this weight restriction, 
the dose rate at the surface of the berm is about 100 times higher with iron
shielding than with heavy concrete.  The berm levels using heavy concrete are 8 times
above the 19$^{\prime}$ dirt standard. 

With this degree of shielding and normal operation,
the surface of berm can be classified as a ``Controlled Area'' with minimal occupancy which implies
a dose rate from 0.25 up to 5 mrem/hr \cite{bib:Dose_rate}.

\subsection{Access pit}

\begin{figure}[htb!]
\vspace{-1mm}
\begin{minipage}[t]{0.65\linewidth}
\hspace{30mm}
\includegraphics[width=4.0in,keepaspectratio]{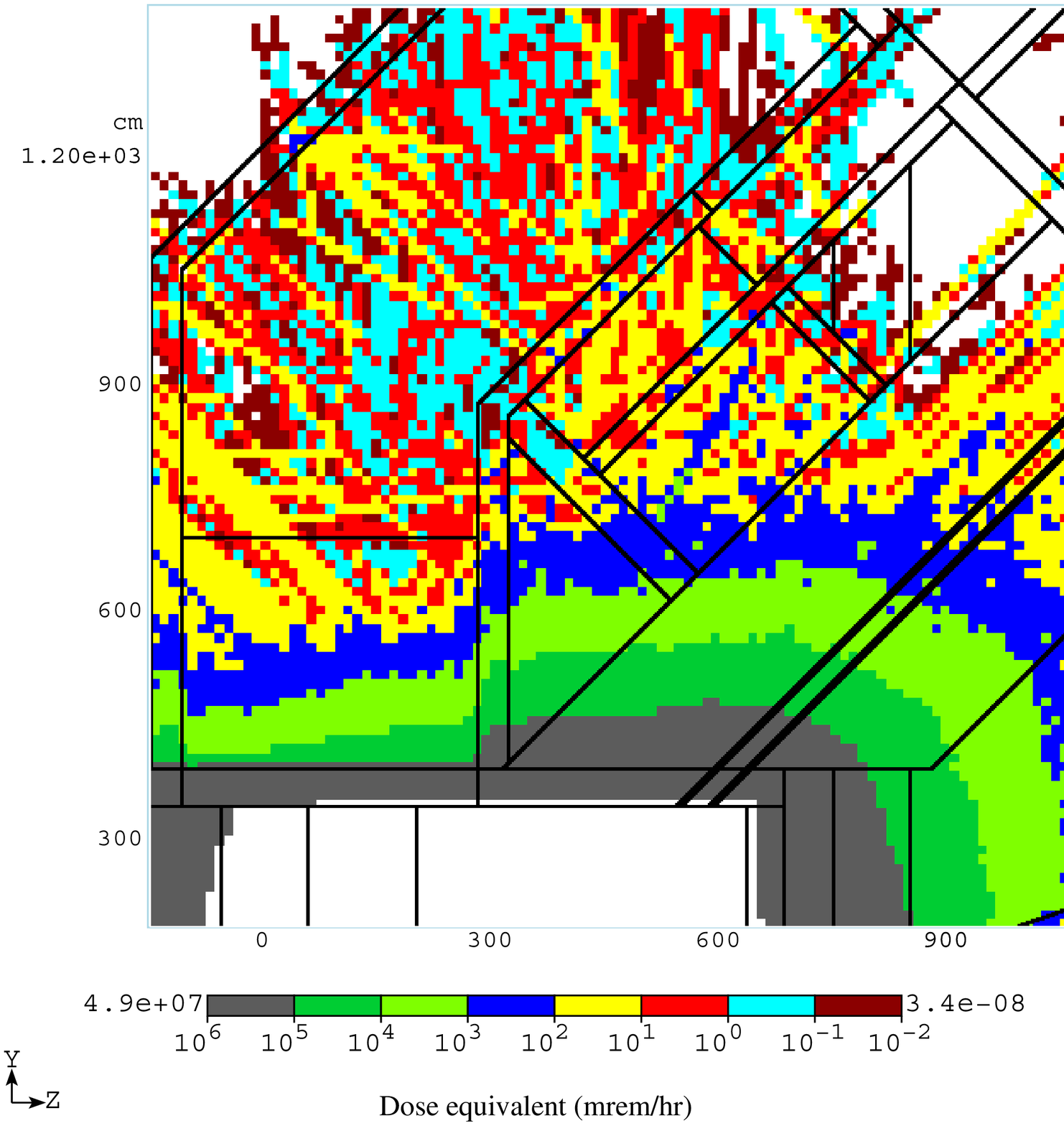}
\end{minipage}

\vspace{8mm}
\begin{minipage}[t]{0.65\linewidth}
\hspace{30mm}
\includegraphics[width=4.0in,keepaspectratio]{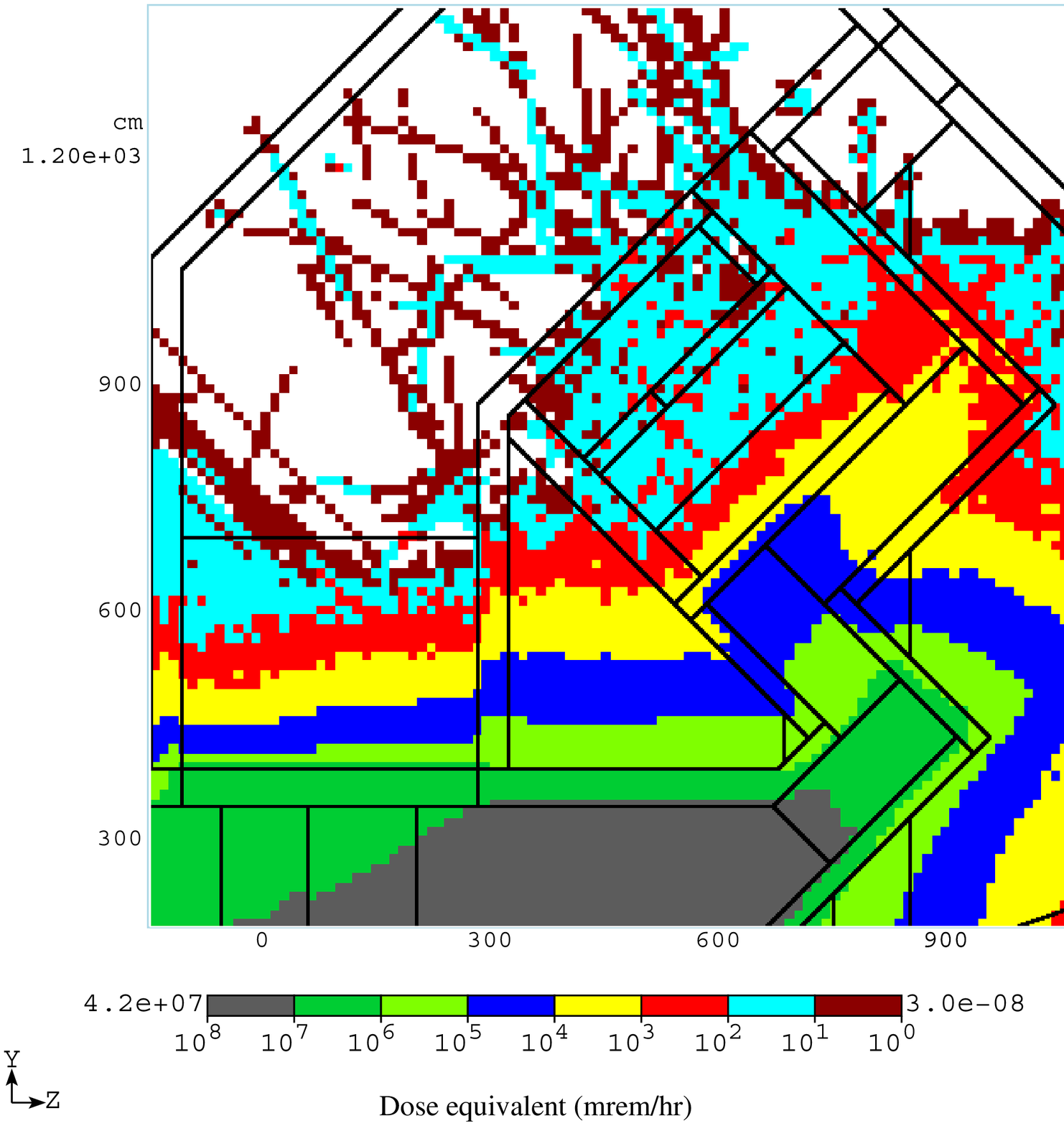}
\end{minipage}

\vspace{7mm}
\caption{Calculated dose distributions in the access pit of the MTA at lower (bottom) and upper (top) level. }
\label{mta_access_pit}
\end{figure}

All the calculations described hereinafter were performed with the 1-cm thick copper disk as a target.
The calculated dose distributions in the access pit are shown in Fig.~\ref{mta_access_pit}.
From the target hall to access pit, a dose reduction  within a factor of $10^6$ is observed,
which means this is a typical thick-shielding problem.
Therefore, using a variance reduction technique like mathematical expectation method is justified and mandatory.

\clearpage

In addition, all the calculations were performed with the MCNP option turned 'on' to provide the most accurate
available at present treatment of low-energy (under 20 MeV) neutron transport \cite{bib:MCNP}. The MCNP option
is essential to get reliable results because such neutrons dominate in the target hall
(see next section).

At both lower and upper level one can see a number of hot spots with dose level from 10 to 100 mrem/hr. An examination
of the calculated two-dimensional dose distribution gives rise to a conclusion that the main weakness of the existing
shielding is in significant amount of empty space in the labyrinth (see Fig.~\ref{mta_lower_level}).
A proper choice of a local shielding in the labyrinth would help to reduce the dose 
at the lower level of the access pit.

During normal operation the access pit is expected to be classified as a ``Radiation Area'' consisting of
rigid barriers with locked
gates (requirements for a dose rate from 5 up to 100 mrem/hr \cite{bib:Dose_rate}).

\subsection{Service building and parking lot}

\begin{figure}[htb!]
\vspace{2mm}
\centering\epsfig{figure=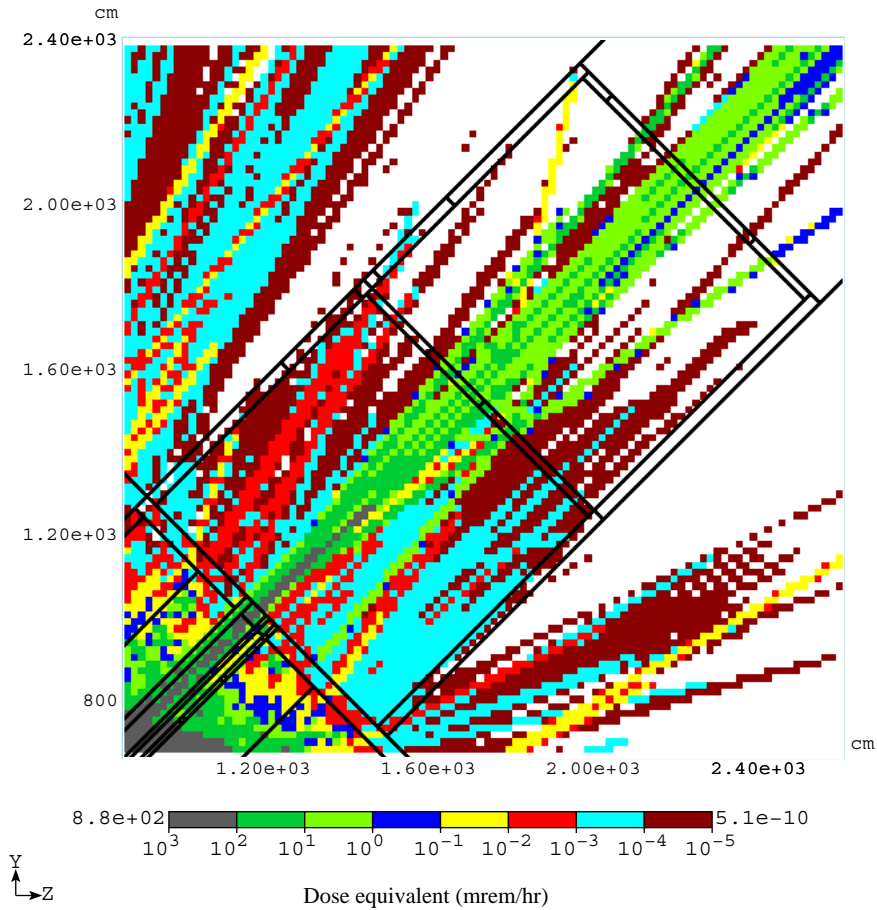,width=0.75\linewidth}
\vspace{6mm}
\caption
{Calculated dose distribution in the service building of the MTA.}
\label{mta_cryo_room_dose}
\end{figure}

\noindent
The dose distribution in and surrounding the service building is computed
for the region of $90$ cm $\leq X \leq 260$ cm, and presented in
Fig.~\ref{mta_cryo_room_dose}. There are a number of hot spots in parking lot with
dose levels ranging from 10 to 100 mrem/hr. Inside the refrigerator room and near the penetrations
the dose is from 100 to 1000 mrem/hr.
One can see also that the highest contribution
to the dose comes from the 10$^{\prime \prime}$ penetration which is in between
the 4$^{\prime \prime}$ and 8$^{\prime \prime}$ ones but at a different height in $X$.

\begin{figure}[htb!]
\vspace{-8mm}
\noindent
\hspace{-5mm}
\begin{minipage}[t]{0.53\linewidth}
\centering\epsfig{figure=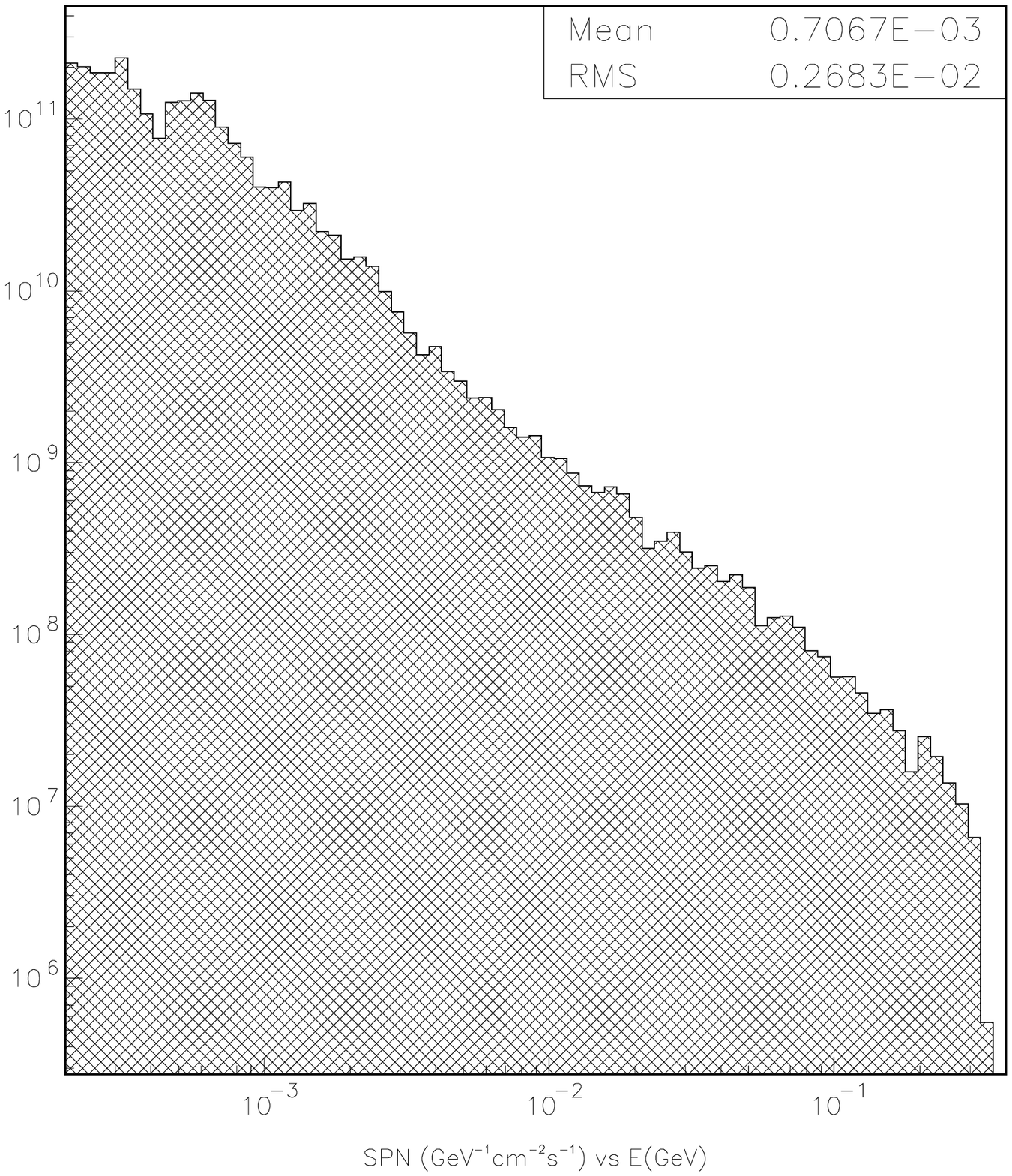,width=\linewidth}
\end{minipage}
\hfill
\hspace{-4mm}
\begin{minipage}[t]{0.53\linewidth}
\centering\epsfig{figure=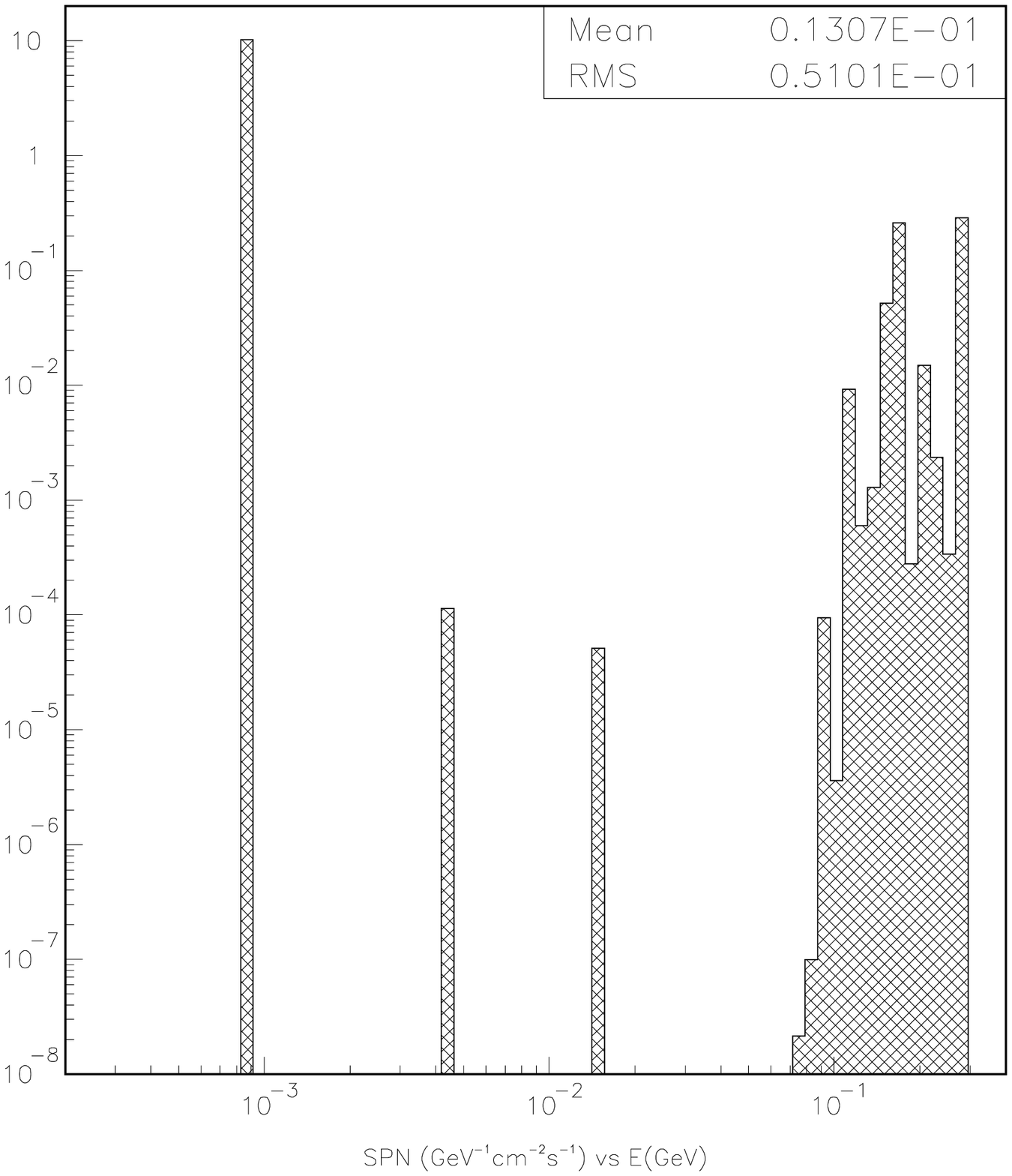,width=\linewidth}
\end{minipage}

\vspace{-7mm}
\caption{Calculated neutron spectra inside the 10$^{\prime \prime}$ penetration at its end near target hall (left)
and refrigerator room (right). }
\label{mta_neutron_spectra}
\end{figure}

Three of the six penetrations are
shown in the Figure: one 8$^{\prime \prime}$ and two 4$^{\prime
\prime}$ in diameter. The most distinctive feature of the
distribution is existence of a directed, intense neutron beam shooting through the
penetrations. The dirt between the target hall and service building
serves, in fact, as a collimator for neutrons generated in proton
collisions with target nuclei \cite{Kamran}.  As a result, in the
service building and at parking lot one has a well collimated, low divergence beam
composed of high energy ($\sim$200 MeV) neutrons. Neutron spectra at both ends of the
largest (10$^{\prime \prime}$) penetration are shown in Fig.~\ref{mta_neutron_spectra}.

\noindent
To reduce the dose in the parking lot and in the service building, the following options were examined:

\begin{itemize}
    \item A wall in the target hall in front of the penetrations.
    \item A wall instead of the door between the two rooms in the service building.
    \item Two iron collimators, 2$^{\prime \prime}$ thick and 20$^{\prime \prime}$ in length,
     placed inside the 10$^{\prime \prime}$ penetration (at both ends) as well as
          two iron collimators, 1$^{\prime \prime}$ thick and 20$^{\prime \prime}$ in length,
     placed inside the 8$^{\prime \prime}$ penetration (at both ends).
\end{itemize}

\clearpage

\begin{figure}[htb!]
\vspace{-1mm}
\noindent
\hspace{-5mm}
\begin{minipage}[t]{0.54\linewidth}
\centering\epsfig{figure=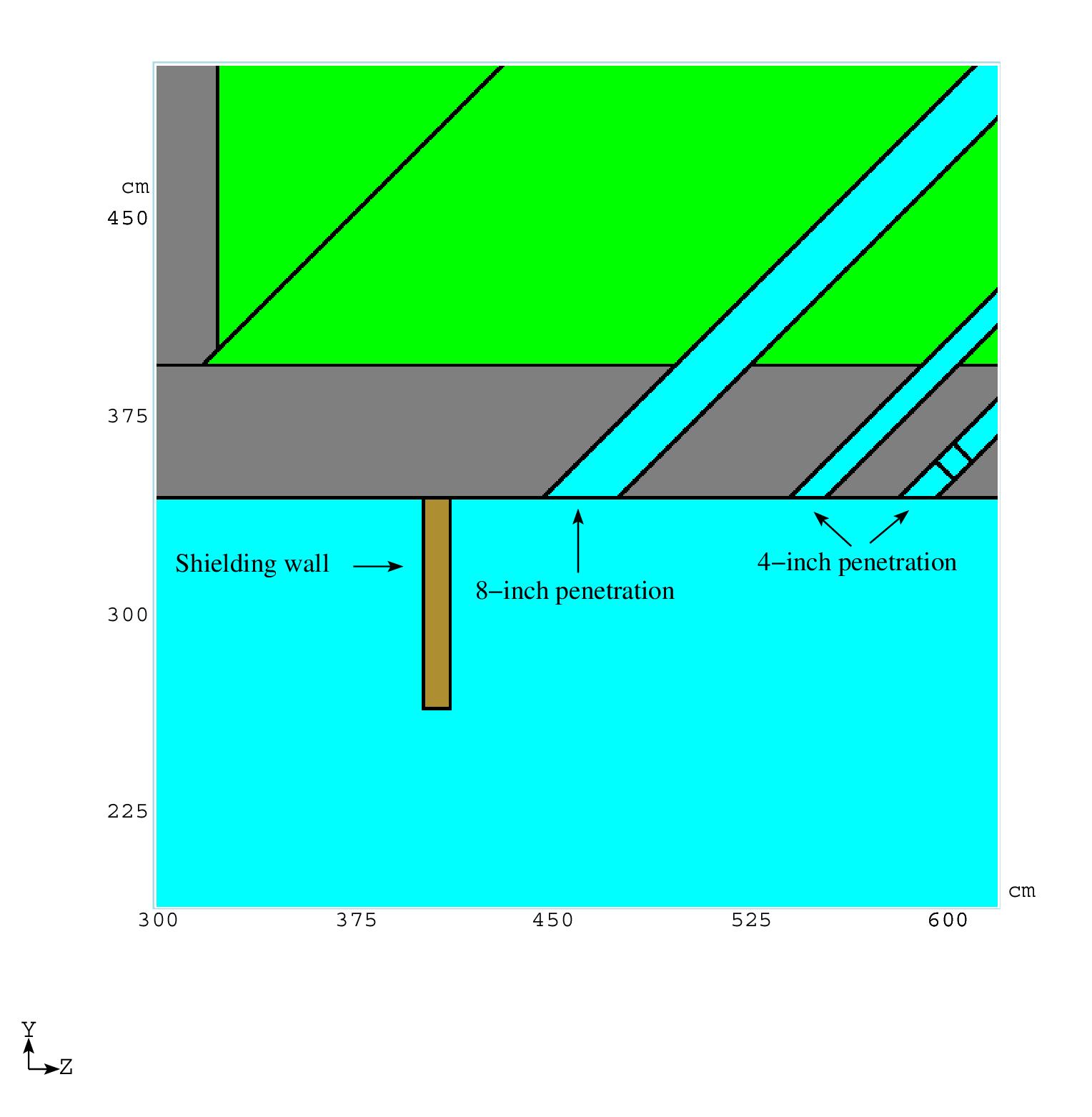,width=\linewidth}
\end{minipage}
\hfill
\hspace{-4mm}
\begin{minipage}[t]{0.54\linewidth}
\centering\epsfig{figure=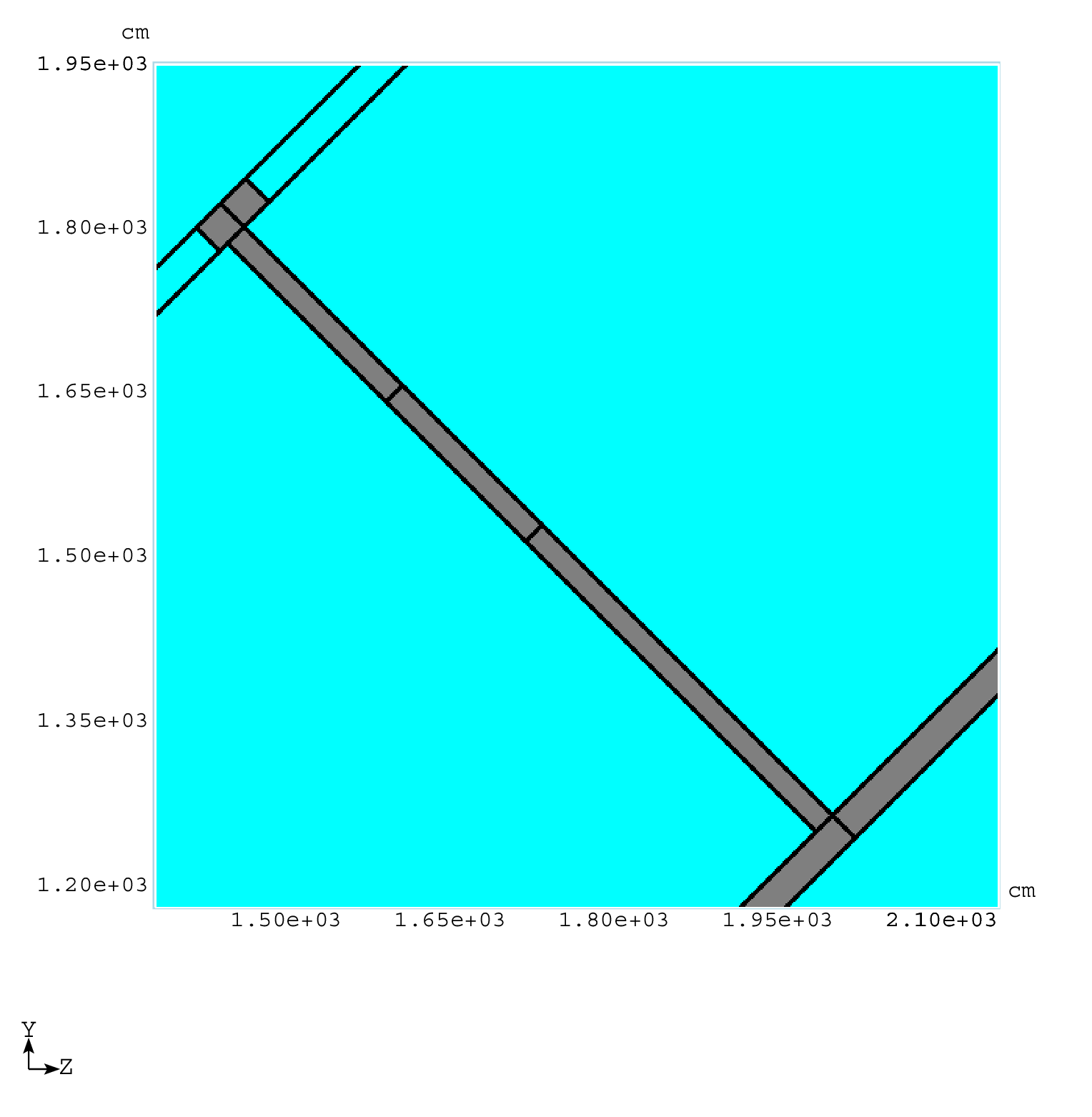,width=\linewidth}
\end{minipage}

\vspace{-4mm}
\hspace{-5mm}
\begin{minipage}[t]{0.54\linewidth}
\centering\epsfig{figure=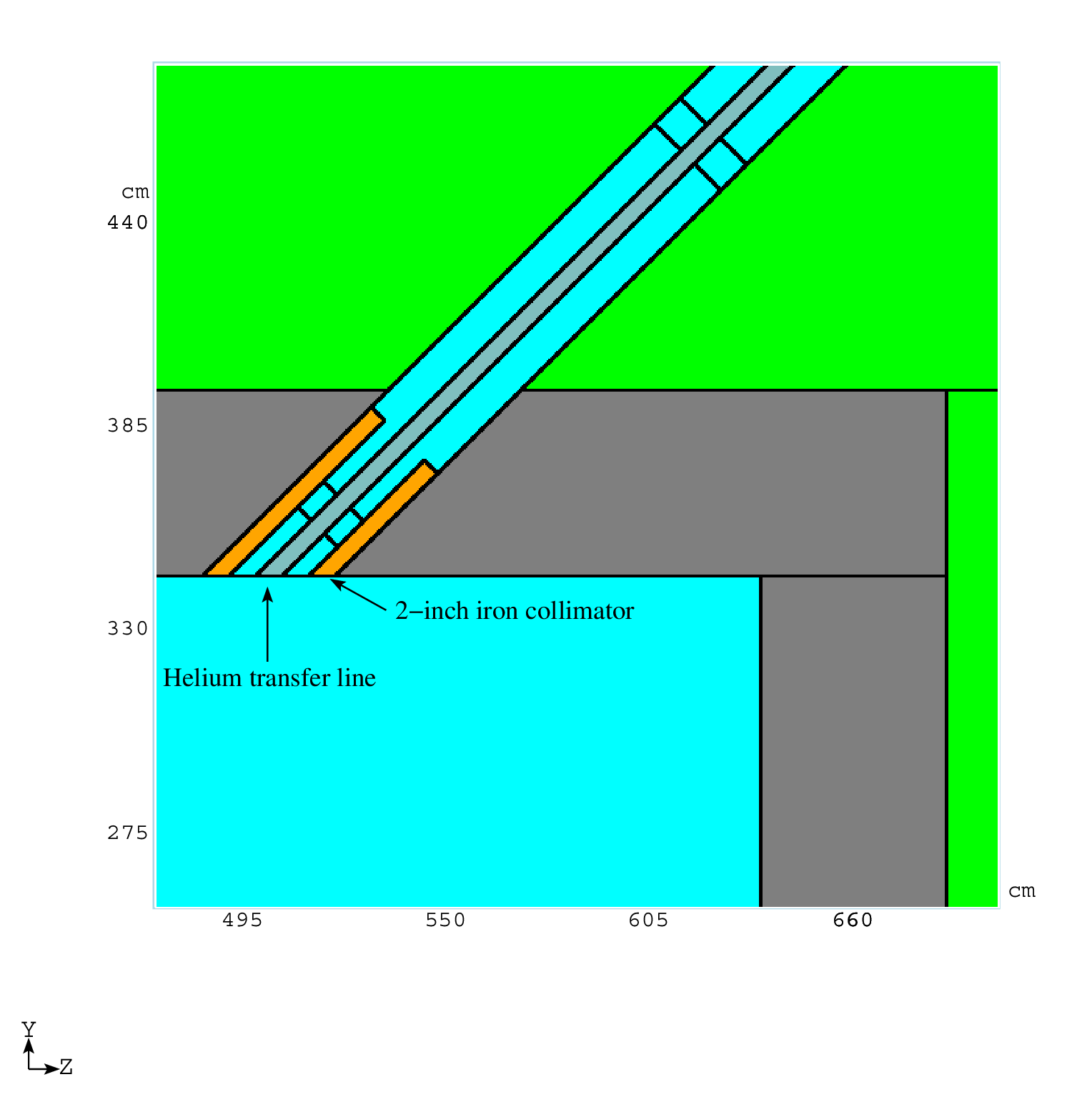,width=\linewidth}
\end{minipage}
\hfill
\hspace{-4mm}
\begin{minipage}[t]{0.54\linewidth}
\centering\epsfig{figure=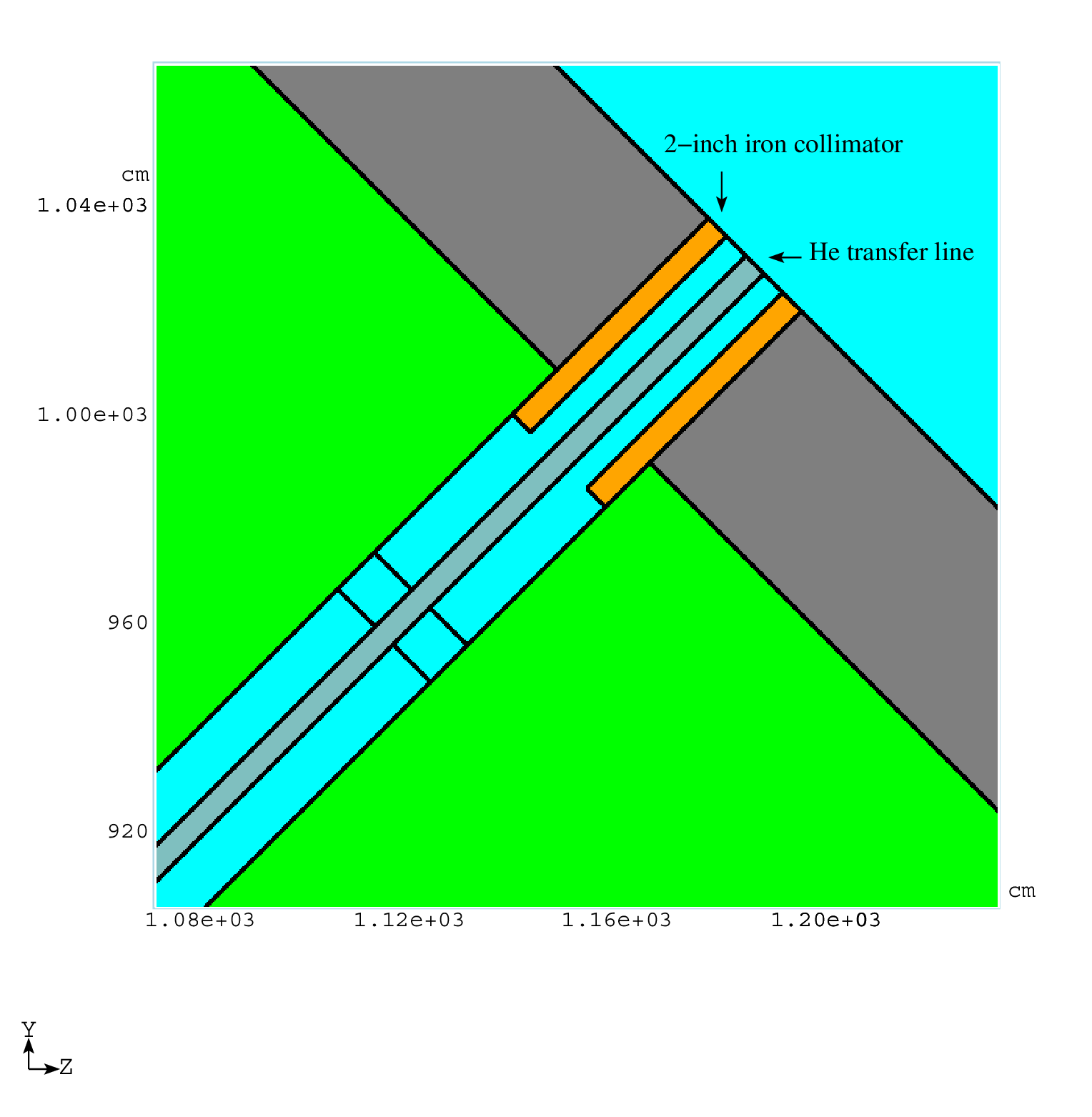,width=\linewidth}
\end{minipage}

\vspace{-3mm}
\caption{Three options for additional shielding: (i) a wall in the target hall (top, left);
(ii) a wall instead of the inner door in the service building (top, right);
(iii) iron collimators in the 8$^{\prime \prime}$ and 10$^{\prime \prime}$ penetrations at both ends (bottom, left and right). }
\label{add_shield_options}
\end{figure}

\begin{figure}[htb!]
\vspace{-1mm}
\noindent
\hspace{-5mm}
\begin{minipage}[t]{0.52\linewidth}
\centering\epsfig{figure=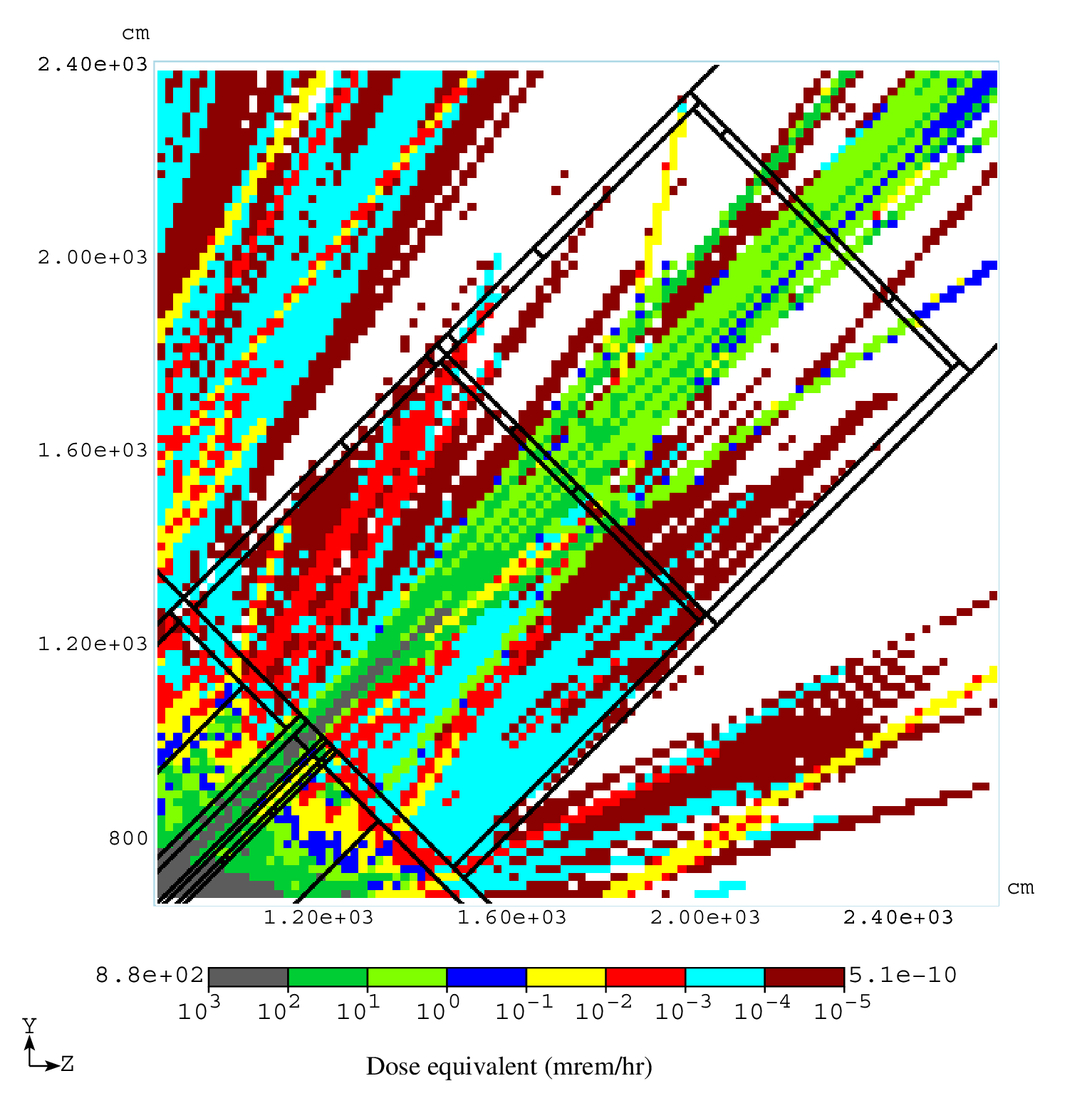,width=\linewidth}
\end{minipage}
\hfill
\hspace{-4mm}
\begin{minipage}[t]{0.52\linewidth}
\centering\epsfig{figure=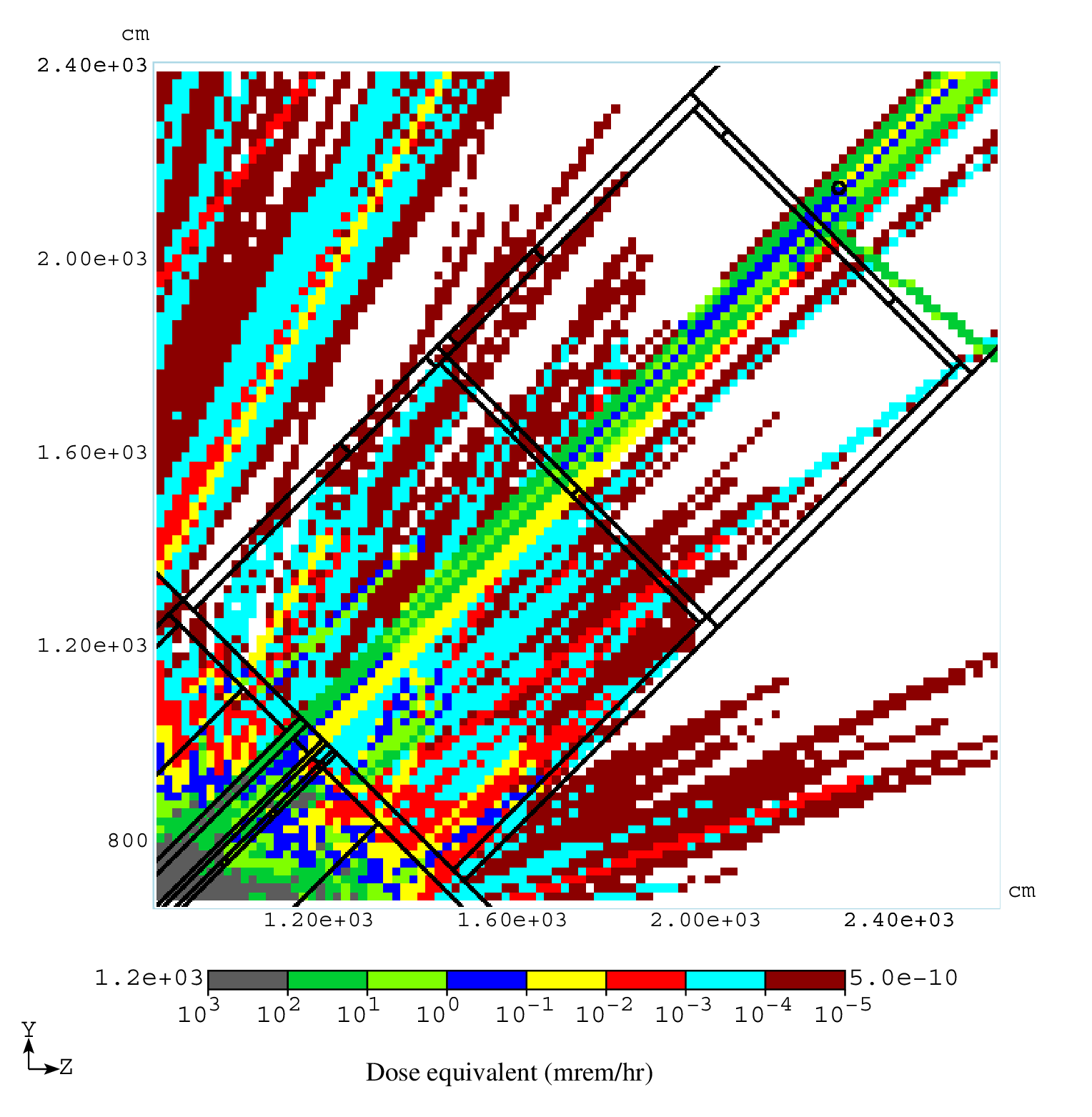,width=\linewidth}
\end{minipage}

\vspace{5mm}
\hspace{-5mm}
\begin{minipage}[t]{0.52\linewidth}
\centering\epsfig{figure=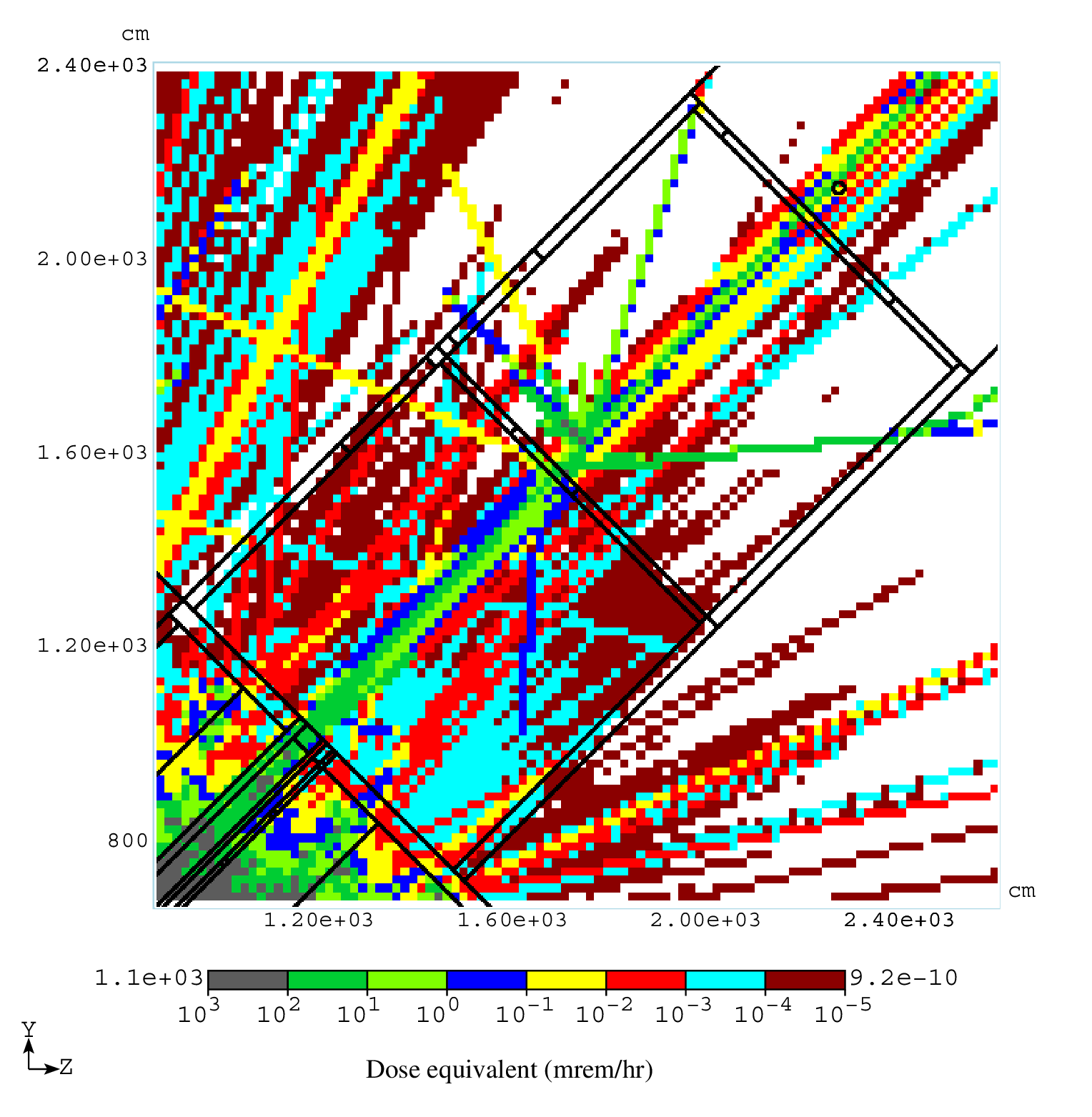,width=\linewidth}
\end{minipage}
\hfill
\hspace{-4mm}
\begin{minipage}[t]{0.52\linewidth}
\centering\epsfig{figure=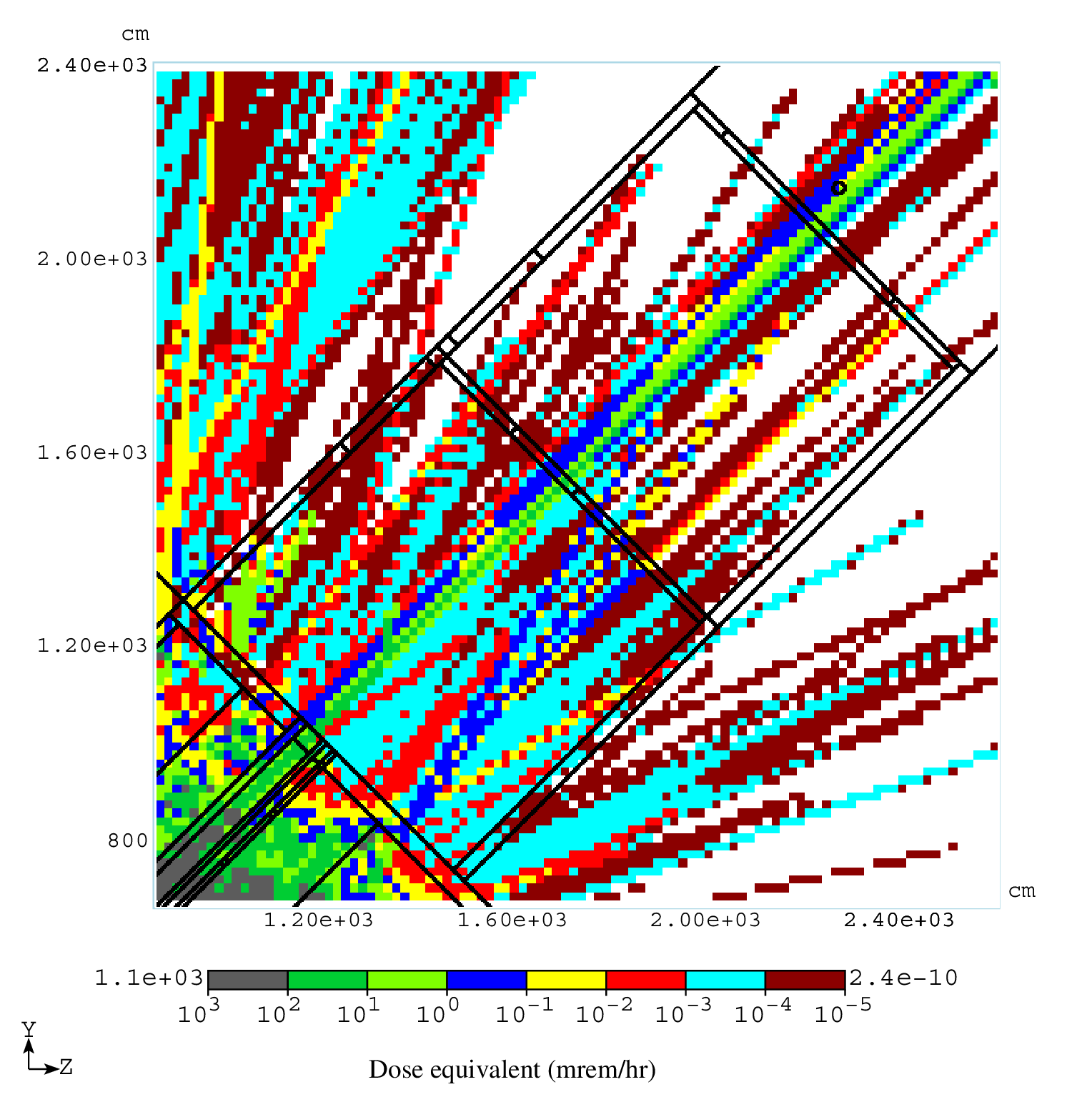,width=\linewidth}
\end{minipage}

\vspace{-1mm}
\caption{Calculated dose distributions in the refrigerator and compressor rooms as well as at parking lot
for the following shielding options: (i) unshielded penetrations (top, left); (ii) a shielding wall in the target hall
in front of the penetrations (top, right); (iii) a 20-cm thick concrete wall instead of the door between the
refrigerator and compressor room (bottom, left); (iv) 5-cm thick and 50-cm long iron collimators at both ends of the
25-cm penetration (bottom, right).}
\label{mta_cryo_room}
\end{figure}

The three options for additional shielding are shown in Figure~\ref{add_shield_options}.
The wall in the target hall was considered consisting
of two parts: a concrete pedestal (lower) and tungsten shielding itself (upper). Thickness and position of the wall
were not optimized.    The tungsten was chosen because of
its high material density and absence of pronounced magnetic properties. The latter is important from the standpoint
of mechanical stability in the presence of superconducting magnets (in the vicinity of the wall) in the event of a quench.  
As for the third option, thicknesses
of the iron collimators were chosen to fit the remaining empty space in the 8$^{\prime \prime}$ and 10$^{\prime \prime}$
penetrations with the helium transfer lines (6$^{\prime \prime}$ in diameter) in place.
The dose distributions in the service building, calculated for the three options and compared with the
initial distribution for unshielded penetrations, are shown in Fig.~\ref{mta_cryo_room}.

\begin{figure}[htb!]
\vspace{-1mm}
\centering\epsfig{figure=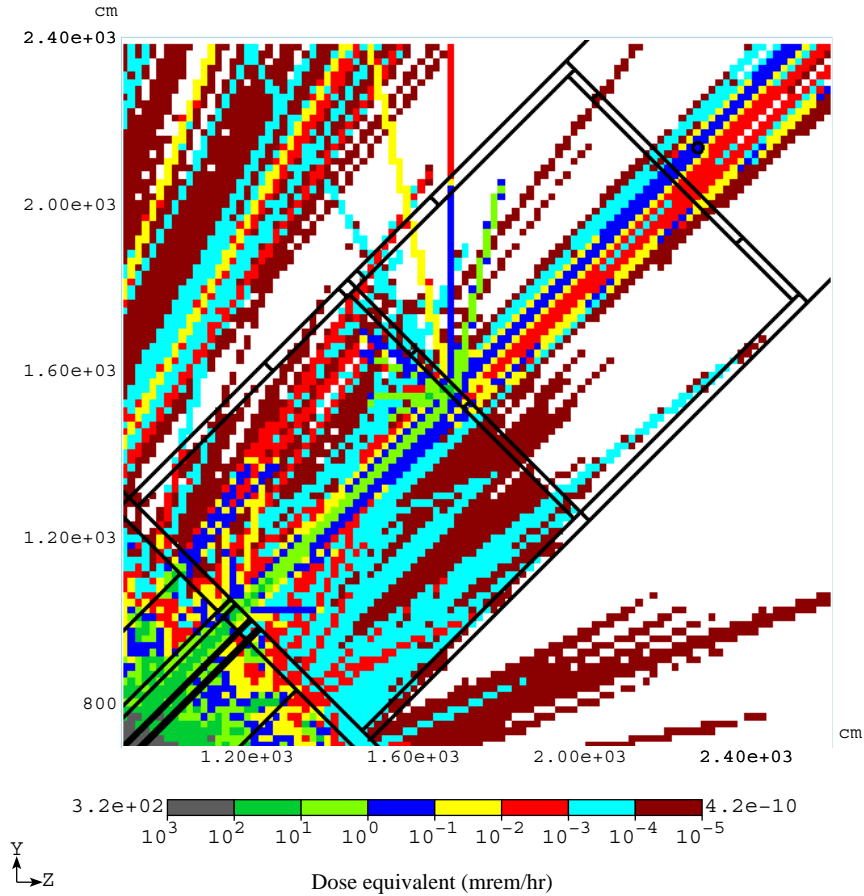,width=0.8\linewidth}
\vspace{-2mm}
\caption
{Calculated dose distribution in the service building and around at normal operation for a combination 
of the second and third shielding options (normalization is per $10^{14}$ protons per second).}
\label{mta_cryo_room_final}
\end{figure}

One can see from the qualitative comparison that both
the second and third option provide better shielding and give rise to a lower dose level at parking lot
when compared to the first option. Therefore, a separate calculation was performed for a combination of
the second and third options, \emph{i.e.} with both the wall instead of the inner door in the service building
and iron collimators installed. The calculated dose distribution is shown in Fig.~\ref{mta_cryo_room_final}. 
One can see that the combined shielding reduces the dose at the parking lot to 0.1--1 mrem/hr.

\noindent
In addition, integral dose was calculated for a cylindrical tissue-equivalent model of a human body placed at the parking lot
near the roll-up
door to the compressor room, \emph{i.e.} at the hottest place. The predicted integral dose equals to 0.4 mrem/hr, but
with a high
statistical uncertainty (1$\sigma \simeq 50\%$).

During normal operation, the public parking lot must be considered as a normal (not
controlled) area without postings. Therefore, the dose level
there must not exceed 0.05 mrem/hr~\cite{bib:Dose_rate}. To
satisfy this requirement, an additional removable 50-cm concrete shielding block
is necessary near the internal wall in the refrigerator room.
At the same time, the compressor and refrigerator
rooms are expected to be classified as ``Controlled Area'' with minimal
occupancy (a dose rate from 0.25 up to 5 mrem/hr) and ``Radiation Area'' with
rigid barriers with locked gates (a rate from 5 up 100 mrem/hr), respectively
\cite{bib:Dose_rate}.

\section{Conclusions}

Within the framework of a credible accident scenario, a beam accident
at the MuCool Test Area is less severe than normal operation.  It is
the normal operating conditions that determine the level of shielding required.

Further, it has been shown that shielding sandwiches of high-density concrete and
dirt provide a much improved dose attenuation above the MTA than
iron-dirt sandwiches considering the load capacity of the hall enclosure. 
Since the enclosure ceiling only supports 19 equivalent feet of compacted soil,
only 10.5$^{\prime}$ of heavy concrete and 5$^{\prime}$ of iron can be supported,
assuming 2$^{\prime}$ of berm is required on top. The corresponding dose rates
for full Linac intensity range from 1 to 5 mrem/hr for the considered targets 
(liquid hydrogen and copper, respectively)
and heavy concrete shielding blocks. The rates for iron shielding range from 100 to 500 mrem/hr.
The heavy concrete alternative also allows the total shielding height to be reduced from
the 16.4$^{\prime}$ pure-dirt height to at least 12.5$^{\prime}$. 
Overall, the heavy concrete-dirt shielding is preferable.

A solid concrete wall replacing the inner door in the service building,
iron collimators situated inside the 8$^{\prime \prime}$ and 10$^{\prime
\prime}$ penetrations as well as additional 50-cm concrete block near the inner wall
in the refrigerator room  are required to suppress the outgoing high-energy neutron beam
to a predicted dose level at the parking lot not exceeding 0.05 mrem/hr for
a beam intensity of $10^{14}$ protons per second. Sand could replace the iron collimators,
if used along the entire length of the penetrations.

After implementing all of the shielding described above, normal operation requires the
different areas around the MTA target hall to be classified as follows \cite{bib:Dose_rate}:

\begin{itemize}
    \item Berm above the target hall -- \textit{Controlled Area} of minimal occupancy (0.25 - 5 mrem/hr).
    \item Access pit -- \textbf{Radiation Area} with rigid barriers with locked gates (5 - 100 mrem/hr).
    \item Refrigerator room -- \textbf{Radiation Area} with rigid barriers with locked gates (5 - 100 mrem/hr).
    \item Compressor room -- \textit{Controlled Area} of minimal occupancy (0.25 - 5 mrem/hr).
    \item Parking lot -- Unlimited occupancy area without any precautions (dose rate below 0.05 mrem/hr).
\end{itemize}

\section{Acknowledgements}

The authors are thankful to Don Cossairt, Kamran Vaziri, Michael Gerardi, Bill Higgins, and Nikolai Mokhov
of Fermilab for helpful discussions.

\vspace{3mm}
\noindent
The work was supported by the Illinois
Board of Higher Education with the Higher Education Cooperative
Act Grant and Universities Research Association, Inc., under
contract DE-AC02-76CH03000 with the U. S. Department of Energy.


\begin{thebibliography}{9}

\bibitem{bib:MuCool} \underline{http://www.fnal.gov/projects/muon\_collider/cool/cool.html}; D.~Errede, R.~Alber,
   A.~Bross \emph{et al.}
   Proc. of the 2003 Part. Accel. Conf., Portland, OR, USA, May 2003.

\bibitem{bib:mars} N.V.~Mokhov, ``The MARS Code System User's Guide'', Fermilab-FN-628
   (1995);  N.~V.~Mokhov, O.~E.~Krivosheev, ``MARS Code Status'', Proc. of the
   Monte Carlo 2000 Conference, Lisbon, October 23-26, 2000, Springer, p.~943;
   Fermilab-Conf-00/181 (2000); \underline{http://www-ap.fnal.gov/MARS/}

\bibitem{Mokhov} N. Mokhov, Private communication, April 2004.

\bibitem{bib:dose_attenuation_curve} The Proton Driver Design Study, Fermilab-TM-2136, Chapter 10.4.4, December 2000.

\bibitem{bib:Dose_rate} ``Fermilab Radiological Control Manual'', Article 236, \underline{http://www-esh.fnal.gov/FRCM/}.

\bibitem{Kamran} K. Vaziri, ``Dose Attenuation Methodology for
NuMI Labyrinths, Penetrations and Tunnels``, Fermilab RP Note 140,
May 2003; K. Vaziri, Private communication, November 2003.

\bibitem{bib:MCNP} J.F. Briesmeister, editor, ``MCNP - A General Monte Carlo N-Particle Transport Code'', Version 4C.
        Pub. LA-13709-M, Los-Alamos National Laboratory (2000).

\end{thebibliography}
\end{document}